\documentclass[aps,pre,twocolumn,preprintnumbers,showpacs,superscriptaddress]{revtex4}

\usepackage{graphicx}

\begin{document}

\title{Naming Games in Two-Dimensional and Small-World-Connected Random Geometric Networks}

\author{Qiming Lu}
\email{luq2@rpi.edu}
\affiliation{Department of Physics, Applied Physics, and Astronomy,
Rensselaer Polytechnic Institute, 110 8th Street, Troy, NY 12180-3590, USA}
\affiliation{Center for Pervasive Computing and Networking,
Rensselaer Polytechnic Institute, 110 8th Street, Troy, NY 12180-3590, USA}

\author{G. Korniss}
\email{korniss@rpi.edu}
\affiliation{Department of Physics, Applied Physics, and Astronomy,
Rensselaer Polytechnic Institute, 110 8th Street, Troy, NY 12180-3590, USA}
\affiliation{Center for Pervasive Computing and Networking,
Rensselaer Polytechnic Institute, 110 8th Street, Troy, NY 12180-3590, USA}

\author{B. K. Szymanski}
\email{szymab@rpi.edu}
\affiliation{Department of Computer Science,
Rensselaer Polytechnic Institute, 110 8th Street, Troy, NY 12180-3590, USA}
\affiliation{Center for Pervasive Computing and Networking,
Rensselaer Polytechnic Institute, 110 8th Street, Troy, NY 12180-3590, USA}

\begin{abstract}
We investigate a prototypical agent-based model, the Naming Game, on
two-dimensional random geometric networks. The Naming Game [A.
Baronchelli et al., J. Stat. Mech.: Theory Exp. (2006) P06014.] is a
minimal model, employing local communications that captures the
emergence of shared communication schemes (languages) in a
population of autonomous semiotic agents. Implementing the Naming
Games with local broadcasts on random geometric graphs, serves as a
model for agreement dynamics in large-scale, autonomously operating
wireless sensor networks. Further, it captures essential features of
the scaling properties of the agreement process for
spatially-embedded autonomous agents.
Among the relevant observables capturing the temporal properties of the agreement process,
we investigate the cluster-size distribution and the distribution of the agreement times,
both exhibiting dynamic scaling.
We also present results for the case when a small density of long-range communication links are
added on top of the random geometric graph, resulting in a
``small-world"-like network and yielding a significantly reduced
time to reach global agreement.
We construct a finite-size scaling analysis for the agreement times in this case.
\end{abstract}

\pacs{89.75.Fb, 
      05.65.+b 
      }

\date{November 21, 2007}

\maketitle

\section{Introduction}

Reaching agreement without global coordination is of fundamental
interest in large-scale autonomous multi-agent systems. In the
context of social systems, the objective is to understand and
predict the emergence of large-scale population-level patterns
arising from empirically supported local interaction rules between
individuals (e.g., humans). Examples for such phenomena driven by
social dynamics include the emergence and the evolution of
languages \cite{Nowak_JTB1999,Nowak_Science2001,Nowak_PNAS2004} or
opinion formation
\cite{Castellano2005,Eli,Deffuant,Durlauf,KR2003,Sznajd,Kozma2007,Benczik2007,Antal2005}.

The creation of shared classification schemes in a system of
artificial and networked autonomous agents can also be relevant from
a system-design viewpoint, e.g., for sensor networks
\cite{Lee_2005,Collier_2004}. Envision a scenario where mobile or
static sensor nodes are deployed in a large spatially-extended
region and the environment is unknown, possibly hostile, the tasks
are unforeseeable, and the sensor nodes have no prior classification
scheme/language to communicate regarding detecting and sensing
objects. Since subsequent efficient operation of the sensor network
inherently relies on unique object identification, the autonomous
development of a common ``language" for all nodes is crucial at the
exploration stage after network deployment.

To this end, in this paper we consider and slightly modify a simple
set of rules, referred to as Language or Naming Games (NG),
originally proposed in the context of semiotic dynamics
\cite{Steels,Kirby}. Such problems have become of technological
interest to study how artificial agents or robots can invent common
classification or tagging schemes from scratch without human
intervention \cite{Steels,Kirby}. The original model
\cite{Steels,Steels_1998,Steels_1995,Steels_1997} was constructed to
account for the emergence of shared vocabularies or conventions in a
community of interacting agents. More recently, a simplified version
of the NG was proposed and studied on various network topologies by
Baronchelli et al.
\cite{Baronchelli_2005a,Baronchelli_2005b,Baronchelli_2006a}, and by
Dall'Asta et al. \cite{Baronchelli_2006b,Baronchelli_2006c} The
advantage of studying a minimal model is that one can gain a deeper
understanding of the spontaneous self-organization process of
networked autonomous agents in the context of reaching global
agreement, and can extract quantitative scaling properties for
systems with a large number of agents.

In the context of artificial agents, there are other possible
scenarios when the NG algorithm, in addition to being interesting in
its own merit in studying agreement dynamics on various networks,
can also be particularly useful from a system-design viewpoint. That
can be the case when one does not intend the outcome of the
agreement process among many agents to be easily predictable. The
actual process of electing a ``leader" or coordinator among sensor
nodes may actually be such a scenario. The leader must typically be
a trusted node, with possible responsibilities ranging from routing
coordination to key distribution \cite{DeCleene}. Standard leader
election (LE) algorithms \cite{Angluin,HS,LeLann,MWV,VDIKT} are
essentially based on finding global extremum (e.g., maximum) through
local communications \cite{Angluin,HS,LeLann}. Thus, the elections
can be stolen by placing a node in the network with a sufficiently
high ID (e.g., the largest number allowed by the number
representation scheme of the sensor chips.) Along these lines, a
possible application of the NG algorithm is autonomous key creation
or selection for encrypted communication in a community of sensor
nodes. Instead of having a centralized or hierarchial key management
system with domain and area key distributors \cite{DeCleene}, group
of sensor nodes can elect a key distributor or a security key for
secure communications between group members.

This work is an expanded version of our preliminary results
\cite{Lu_AAAI2006}. In addition to showing more detailed and
extended simulation results, we also study and analyze different
aspects of the behavior of the model, in particular, the probability
distribution of the agreement times and the cluster-size
distribution in the NG on random geometrical graphs (RGGs). Further,
we construct and present finite-size scaling for the agreement times
in Small-World \cite{Watts98} (SW)-connected RGGs. The remainder of
the paper is organized as follows. In Sec.~II we briefly review
recent results on the NG on various regular and complex networks. In
Sec.~III we define and present results on the NG with {\em local
broadcast} on RGGs, motivated by communication protocols in sensor
networks. In Sec.~IV we present and discuss results on the NG on
SW-connected RGGs. Section V concludes our paper with a brief
summary and outlook.

\section{Background and Prior Results on the Naming Game}

In the simplified version of the NG, agents perform {\em pairwise}
games in order to reach agreement on the name to assign to a {\em
single} object. This version of the NG was investigated on
fully-connected (FC) (also referred to as mean-field or homogeneous
mixing) \cite{Baronchelli_2005a,Baronchelli_2005b}, on regular
\cite{Baronchelli_2006a}, on small-world (SW)
\cite{Baronchelli_2006b,Lin2006}, and on scale-free networks
\cite{Baronchelli_2006c,Baronchelli_2006d}. In the FC network, each
agent has a chance to meet with all others and compare their current
local vocabularies (list of ``synonyms") before updating them. On
regular networks, agents have only a limited and fixed number of
neighbors on a one-, two-, etc., dimensional grid with whom they can
interact/communicate. The communication in both cases is ``local",
in that {\em pairs of agents} are selected to interact and to update
their vocabularies. The basic algorithmic rules of the NG are as
follows \cite{Baronchelli_2005a,Baronchelli_2006a}. A pair of
neighboring nodes (as defined by the underlying communication
topology), a ``speaker" and a ``listener", are chosen at random
\cite{NG_order}. The speaker will transmit a word from her list of
synonyms to the listener. If the listener has this word, the
communication is termed ``successful", and both players delete all
other words, i.e., collapse their list of synonyms to this one word.
If the listener does not have the word transmitted by the speaker,
she adds it to her list of synonyms without any deletion.

Among the above rules, the restriction to a single object
\cite{Baronchelli_2005a,Baronchelli_2005b} strongly reduces the
complexity of the model, compared to a more general case where the
naming process of multiple objects can be performed simultaneously.
From a linguistic viewpoint, this rather strong restriction is
equivalent to preventing homonymy, and instead, treating all objects
independently. This strong assumption can be more realistic for a
system of artificial agents, where agents assign random numbers
(e.g., chosen from $2^{31}$ integers) as ``words" to new objects. In
this case, the number of potential words can be far grater than the
number of objects, and the probability that two players invent the
same word for different objects (hence giving rise to homonymy) is
negligible.

It was found that employing the above local rules ({\em pair-wise}
interactions), after some time, the agents vocabularies converge to
a unique vocabulary shared among all agents
\cite{Baronchelli_2005a,Baronchelli_2005b,Baronchelli_2006a,Baronchelli_2006b}.
The major differences between the NG on FC graphs and on regular
low-dimensional grids arise in the scaling of the the memory needed
to develop the common language before convergence occurs, and in the
scaling of the time $t_c$ needed to reach global agreement. (The
memory need in the present context is the typical value of the
largest number of words an agent may posses throughout the evolution
of the game \cite{Baronchelli_2005a,Baronchelli_2006a}.)
In the FC network, the convergence process to global agreement is fast
[$t_c\sim{\cal O}(N^{1/2})$ for $N$ agents], but large memory
[${\cal O}(N^{1/2})$] is needed per agent \cite{Baronchelli_2005a}.
For a regular two-dimensional network (or grid), spontaneous
evolution toward a shared dictionary is slow [$t_c\sim{\cal O}(N)$],
but the memory requirement per agent is much less severe [${\cal
O}(1)$] \cite{Baronchelli_2006a}. When the NG is implemented on
Watts-Strogatz \cite{Watts98} SW networks, the agreement dynamics
performs optimally in the sense that the memory needed is small,
while the convergence process is much faster than on the regular networks
[$t_c\sim{\cal O}(N^{0.4})$, closer to that of the FC network]
\cite{Baronchelli_2006b}.

Sensor networks, which are motivating our study, are both spatial and
random. As a large number of sensor nodes are deployed, e.g.,  from
vehicles or aircrafts, they are essentially scattered randomly
across large spatially-extended regions. In the corresponding
abstract graph, two nodes are connected if they mutually fall within
each others transmission range, depending on the emitting power, the
attenuation function and the required minimum signal to noise ratio.
Random geometric graphs (RGGs), also referred to as spatial
Poisson/Boolean graphs, capturing the above scenario, are a common
and well established starting point to study the structural
properties of sensor network, directly related to coverage,
connectivity, and interference. Further, most structural properties
of these networks are discussed in the literature in the context of
continuum percolation \cite{percolation,penrose,Dall_2002}.

The common design challenge of these networks is to find the
optimal connectivity for the nodes: If the connectivity of the
nodes is too low, the coverage is poor and sporadic. If the node
connectivity is too high, interference effects will dominate and
result in degraded signal reception
\cite{Kumar2000,Kumar2004,phtr_networks,krause04,BCS}. From a
topological viewpoint, these networks are, hence, designed to
``live" somewhere above the percolation threshold. This can be
achieved by adjusting the density of sensor nodes and controlling
the emitting power of the nodes; various power-control schemes
have been studied along these lines \cite{Kumar2000,krause04,BCS}.
In this paper we consider RGGs in two-dimensions above the percolation threshold, as
minimal models for the underlying network communication topology.
Further, we consider RGGs with an added small density of ``random"
long-range links. The resulting structure resembles small-world
(SW) networks \cite{Watts98,NEWMAN_SIAM}, also well studied in the
context of artificial \cite{KNGTR03a,LKS_INSS2006} and social
systems \cite{Watts99,NEWMAN_SIAM}.The focus of this work is to
study the NG algorithm on these spatially-embedded random graphs.

\section{Naming Games on Random Geometric Networks}

\subsection{Random Geometric Graphs}
As mentioned above in the Introduction, first we consider random
geometric graphs in two dimensions
\cite{percolation,penrose,Dall_2002} as the simplest topological
structures capturing the essential features of ad hoc sensor
networks. $N$ nodes are uniformly random distributed in an
$L$$\times$$L$ spatial area. For simplicity we consider identical
radio range $R$ for all nodes. Two nodes are connected if they fall
within each other's range. An important parameter in the resulting
random geometric graph is the average degree $\overline{k}$ (defined
as the average number of neighbors per node),
$\overline{k}$$=$$2K/N$, where $K$ is the total number of links and
$N$ is the number of nodes. In random geometrical networks, there is
a critical value of the average degree, $\overline{k}_c$, above
which the largest connected component of the network becomes
proportional to the total number of nodes (the emergence of the
giant component) \cite{percolation,penrose,Dall_2002}. For
two-dimensional RGGs $\overline{k}_c$$\approx$$4.5$
\cite{Dall_2002}. There is a simple relationship between the average
degree $\overline{k}$, the density of nodes $\rho$$=$$N/L^2$, and
the radio range $R$ of the nodes
\cite{percolation,penrose,Dall_2002},
$\overline{k} =\rho \pi R^2$,
which can be used to control the connectivity (average degree) of
the network.

\subsection{The Naming Game with Local Broadcast}
We consider the Naming Game on random geometrical graphs. In the
original context of the NG, agents try to reach agreement in
finding a unique ``word" for an object observed by them. In one of
the above proposed potential applications, agents try to generate a shared
unique key for encrypted communication. For simplicity, we will
use the term ``word" for the latter as well when describing the
algorithm.

Motivated by communication protocols employed by sensor nodes, we
modify the communication rules to make them applicable for sensor
networks. Instead of pairwise communications, nodes will initiate
{\em broadcast} (to all neighbors) in a continuous-time asynchronous
fashion. In this paper we consider the initial condition when the
``vocabulary" of each node is empty. At every elementary time step,
a node is chosen randomly out of $N$ nodes (mimicking Poisson
asynchrony for large $N$). This node (the ``speaker") will broadcast
a word from her list of ``synonyms"; if her list of synonyms is
empty, the speaker randomly invents a word; if she already has
several synonyms, it randomly chooses one. Her neighbors (the
``listeners") compare their vocabularies with the word transmitted
by the speaker. If a listener has this word, she considers the
communication a success, and she deletes all other words, collapsing
her list of synonyms to this one word. If a listener does not have
the word transmitted by the speaker, she adds it to her list of
synonyms without any deletion. If {\em at least one} listener had
the word transmitted, the speaker considers it (at least a partial)
success, and (somewhat optimistically) collapses her list of
synonyms to this one word. At every step, the ``success" rate $S$ is
defined as the fraction of listeners who were successful (i.e.,
those that had the word transmitted by the speaker). From the above
it is clear that one of the successful listeners, if any,
has to report the outcome of the ``word matching" to the speaker. In
order to achieve that efficiently, in real sensor-network implementations
one can employ the ``lecture hall" algorithm \cite{Chen_2005,Szymanski_2007}.
In this paper time $t$ is given in
units of one ``speaker"-initiated broadcast {\em per node}. The main
difference between the above algorithm and the one in Refs.
\cite{Baronchelli_2005a,Baronchelli_2005b,Baronchelli_2006a,Baronchelli_2006b}
is the {\em broadcast} (instead of pairwise communications) and the
underlying network (RGG in this paper) to capture the essential
features of the NG in sensor networks.

\begin{figure}[t]
\vspace*{2.0truecm}
       \includegraphics{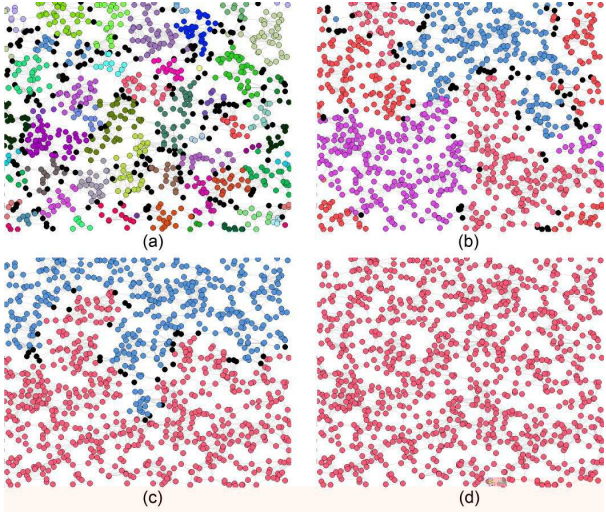}
\vspace*{5.5truecm}
\caption{(Color) Snapshots of the time
evolution of the contents of the agents' word lists during the
process of reaching global agreement on RGG for $N=1,000$ nodes at
time (a) $t=1$; (b) $t=43$; (c) $t=169$; (d) $t=291$. The average
degree is $\overline{k}$$\approx$$12$. Initially, the word lists are
empty for all agents. Time, as through the paper,
is measured in units of ``speaker"-initiated broadcasts {\em per node}.
Different colors correspond
to different words, with black indicating nodes with multiple words.
After the early-time increase in the number of different words in
the systems, small spatial clusters sharing the same word quickly
form, then subsequently ``coarsen" until eventually only one global
cluster prevails.}
\label{fig.snapshots}
\end{figure}

\begin{figure}[tb]
\hspace*{-0.5cm}
\includegraphics[width=3.7in]{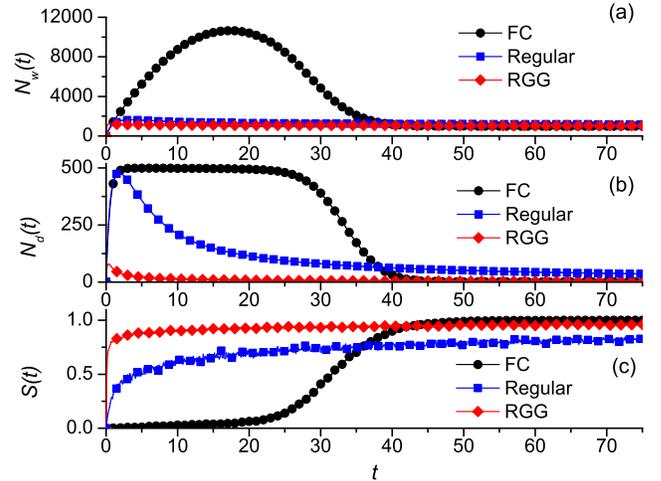}
\vspace*{-1.00truecm}
\caption{(Color online) Time evolution of the relevant
observables in the Naming Game in the fully-connected (FC),
two-dimensional regular (with four nearest neighbors), and random geometric networks
(RGG) for $N$$=$$1024$, averaged over $1,000$ independent
realizations; (a) the total number of words in the
system $N_{w}(t)$; (b) the number of different words $N_{d}(t)$;
(c) the average success rate $S(t)$. The average degree of the
underlying RGG is $\overline{k}$$\approx$$12$. Data for the FC and
$2d$ regular networks are reproduced by our simulations, following
Refs.~\cite{Baronchelli_2005a,Baronchelli_2006a}, for comparison.}
\label{fig.mf-rgg-reg}
\end{figure}

When starting from empty vocabularies, agents invent words randomly.
After time of ${\cal O}(1)$ [on average one speaker-initiated
broadcast per node], ${\cal O}(N/(\overline{k}+1))$ different words
have been created. Following the early-time increase of the number
of different words $N_d(t)$, through local broadcasts, agents slowly
reconcile their ``differences", and eventually will all share the
same word. First, a large number of small spatial clusters sharing
the same word develop. By virtue of the slow coalescence of the
interfaces separating the clusters, more and more of the small
clusters are being eliminated, giving rise to the emergence of
larger clusters, eventually leading to one cluster in which all
nodes are sharing the same word. As suggested by Baronchelli et al.
\cite{Baronchelli_2006a}, this late-time process is analogous to
coarsening, a well-known phenomenon from the theory of domain and
phase ordering in physical and chemical systems \cite{Bray}.
Figure~\ref{fig.snapshots} shows snapshots of vocabularies of the
nodes at different times. For later times, group of nodes which
already share the same word, slowly coarsen, until eventually only
one domain prevails. This behavior is also captured by
Fig.~\ref{fig.mf-rgg-reg}(b), tracing the number of different words
as a function of time $N_d(t)$, eventually reaching global
agreement, $N_d=1$.


\subsection{Basic Scaling Considerations and Analogy with Coarsening}
Before turning to the detailed discussion of our simulation
results, we first sketch the framework of coarsening theory
\cite{Bray}, applicable to the observed late-time dynamics of the
NG on regular $d$-dimensional lattices \cite{Baronchelli_2006a}.
Coarsening has also been observed in other models relevant to
opinion formation and social dynamics \cite{BNFK1996,KR2003}.
Unlike other minimalist (typically two-state) models often
employed to study opinion formation \cite{Durlauf}, such as the
one studied by Sznajd-Weron \& Sznajd \cite{Sznajd}, the
Voter model \cite{intpart,BNFK1996}, or the majority rule model
\cite{KR2003}, in the NG, each agent can be in an {\em unlimited}
number of discrete states (corresponding to a chosen word).
Further, at any instant before reaching global consensus, an agent
can have different possible words for the object.
Because of the potentially unlimited number of discrete states the
agents can assume, the late-stage evolution of the NG resembles that
of infinite-state ($Q$$=$$\infty$) Potts model
\cite{Glazier1990,Grest1983a,Grest1983b,Kaski1985,Kaski1987,Grant1990,Derrida1991,Derrida1996,EBN1997,Majumdar1995a,Majumdar1995b,bAvraham1995,bAvraham2000,Wykes}.

While RGG is a random structure, it is embedded in two dimensions,
and we also attempt to employ elementary scaling arguments from
coarsening theory. According to Ref.~\cite{Baronchelli_2006a}, on
regular $d$-dimensional lattices, the typical size of domains
(each with already agreed upon one word) is governed by a single
length scale $\xi(t)\sim t^{\gamma}$ with $\gamma$$=$$1/2$,
analogous to that of domain formation in systems with a
non-conserved order parameter \cite{Bray}. Thus, in $d$ dimensions
the average domain size $C(t)$ follows
\begin{equation}
C(t) \sim \xi^d(t) \sim t^{d\gamma} \;.
\label{C}
\end{equation}
and the total number of {\em different} words $N_d$ at
time $t$ scales as the typical number of domains
\begin{equation}
N_d(t) \sim \frac{N}{\xi^d(t)}\sim \frac{N}{t^{d\gamma}} \;.
\label{Nd}
\end{equation}
Further, the total number of words $N_w$ ($N_w/N$ being the average memory load per agent),
at this late coarsening
stage, can be written as the number of nodes $N$ plus the number
of nodes with more than one (on average, between one and two) words,
separating the different domains. It is of order of typical number of domains
times the typical length of the interface of one domain, yielding
\begin{equation}
N_w(t)- N \sim \frac{N}{\xi^d(t)}\xi^{d-1}(t) \sim
\frac{N}{\xi(t)} \sim \frac{N}{t^{\gamma}}\;.
\label{Nw}
\end{equation}
Similarly, the ``failure rate" for word matching, $1$$-$$S(t)$,
(where $S(t)$ is the success rate) scales as the fraction of nodes
at the interfaces separating domains with different words
\begin{equation}
1-S(t) \sim \frac{1}{\xi(t)} \sim \frac{1}{t^{\gamma}}\;.
\label{S}
\end{equation}
The main feature of the above power-law decays (up to some
system-size dependent cut-offs) is that the number of different
words $N_d$, the total number of words $N_w$, and the success rate
$S(t)$ only depend on $t$ through the characteristic length scale
$\xi(t)$. Further, for the typical time $t_c$ to reach global
agreement or consensus, one has $\xi^d(t_c)$$\sim$$N$, i.e.,
\begin{equation}
t_c\sim N^{1/(d\gamma)} \;.
\label{tc}
\end{equation}
Unless noted otherwise (as in Sec.~D.1 and 2), our notation, $t_c$,
$N_d(t)$, $N_w(t)$, $C(t)$, and $S(t)$ refer to the {\em
ensemble-averaged} values of these relevant observables.

\subsection{Simulation Results}

Relevant quantities measured in the simulations are the total
number of words in the system $N_{w}(t)$ (corresponding to the
total memory used by the agents for word allocation at time $t$),
the number of different words $N_{d}(t)$, and the average size of
domains/clusters $C(t)$.
Figure~\ref{fig.mf-rgg-reg} displays the time evolution of these
three quantities for the RGG, compared to the fully connected (FC)
and to the $2d$ regular networks. Here, for the comparison, we
reproduced the corresponding data of
Refs.~\cite{Baronchelli_2005a,Baronchelli_2006a}. The behavior of
the NG on RGG is qualitatively very similar to that of the NG on
$2d$ regular graphs. After time of ${\cal O}(1)$, ${\cal
O}(N/(\overline{k}+1))$ different words have been invented
[Fig.~\ref{fig.mf-rgg-reg}(b)].
$N_w(t)$ also reaches its maximum in time of ${\cal O}(1)$
[Fig.~\ref{fig.mf-rgg-reg}(a)].
\begin{figure}[t]
\hspace*{-0.50cm}
\includegraphics[width=3.7in]{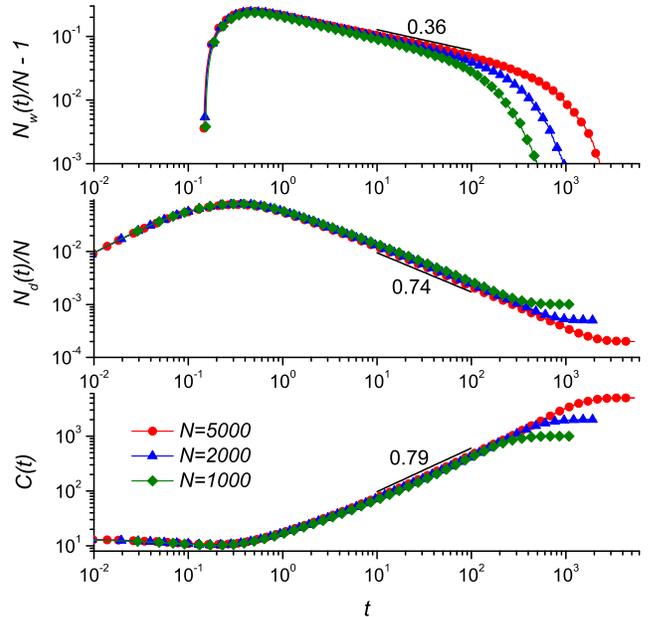}
\vspace*{-1.00truecm}
\caption{(Color online) Time evolution of the relevant
observables in the Naming Game in random geometric networks (RGG)
for three system sizes on log-log scales, averaged over $1,000$
independent realizations; The average degree of the
underlying RGGs is $\overline{k}$$\approx$$12$.(a) the normalized total
number of words in the system $N_{w}(t)/N$$-$$1$; (b) the normalized number
of {\em different} words $N_{d}(t)/N$; (c) the average domain size
$C(t)$. The straight line segments correspond to the best-fit
power-law decays $N_{w}(t)/N$$-$$1$$\sim$$t^{-0.36}$,
$N_d(t)/N$$\sim$$t^{-0.74}$, $C(t)$$\sim$$t^{0.79}$ for
(a), (b), and (c), respectively.}
\label{fig.nd-nw-rgg-scaled}
\end{figure}
Focusing on the late-time behavior of the systems, plotting
$N_w(t)/N$$-$$1$, $N_d(t)/N$, and $C(t)$ vs $t$ on log-log
scales, confirms the power-law decays associated with the
underlying coarsening dynamics, predicted by Eqs.~(\ref{Nw}),
(\ref{Nd}), and (\ref{C}), respectively [Fig.~\ref{fig.nd-nw-rgg-scaled}].
From the data for $C(t)$, we obtain $2\gamma$$=$$0.79\pm0.01$
[Fig.~\ref{fig.nd-nw-rgg-scaled}(c)], while from the data for
$N_d(t)$ and $N_w(t)$, we extract $2\gamma$$=$$0.74\pm0.01$ and
$\gamma$$=$$0.36\pm0.01$, respectively
[Figs.~\ref{fig.nd-nw-rgg-scaled}(b,a)]. Based on our finite-size
results, we can only conclude that the coarsening exponent is in the
range $0.35$$<$$\gamma$$<$$0.40$ for the NG on two dimensional RGG.
Different exponent values extracted from different observables for
{\em finite} systems long hindered the precise determination of the
coarsening exponent in the closely related large-$Q$ Potts model
\cite{Kaski1985,Kaski1987,Grant1990}. There, employing advanced
Monte Carlo renormalization (MCRG) schemes, it was shown that the
coarsening exponent (within error) is $1/2$ \cite{Grant1990}.
However, finite-size effects and very strong transients, in part due
to ``soft domain walls" and domain-wall intersections (``vertices")
can produce values significantly smaller than $1/2$ extracted from
standard MC methods \cite{Kaski1985,Kaski1987,Grant1990}, such as
ours.

\begin{figure}[t]
\hspace*{-0.50cm}
\includegraphics[width=3.7in]{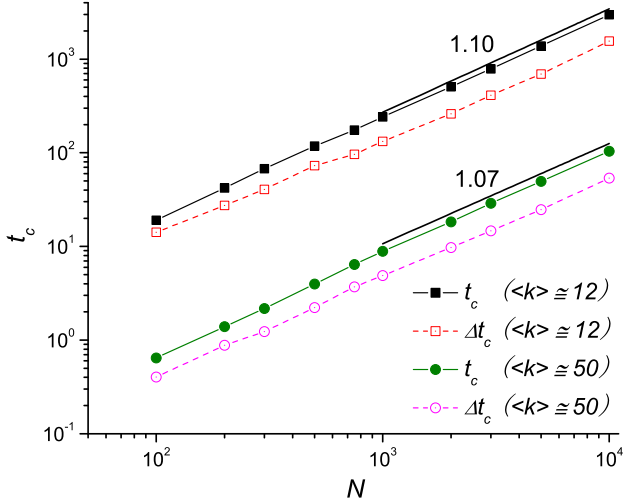}
\vspace*{-1.00truecm}
\caption{(Color online) Average and the standard deviation of the convergence time $t_c$ until
global agreement is reached, as a function of the number of nodes on log-log scales,
averaged over $1,000$ independent realizations of the NG on RGG.
The average degree of the underlying RGGs is $\overline{k}$$\approx$$12$
(squares) and $\overline{k}$$\approx$$50$ (circles).
The straight lines correspond to the best-fit power-laws with exponents $1.10$,
for both the average (solid squares) and the standard deviation (open squares)
of RGGs with $\overline{k}$$\approx$$12$, and $1.07$ for those of RGGs with
$\overline{k}$$\approx$$50$, respectively.}
\label{fig.conv-max-rgg}
\end{figure}
Measuring the time to global agreement, averaged over $1,000$
independent runs (each on a different RGG network realization), we
also obtained the scaling behavior of the agreement time,
$t_c$$\sim$$N^{1.10\pm0.01}$ and $t_c$$\sim$$N^{1.07\pm0.02}$ for
$\overline{k}$$\approx$$12$ and $\overline{k}$$\approx$$50$,
respectively, as shown in Fig.~\ref{fig.conv-max-rgg}. The
corresponding scaling exponents both somewhat deviate from the one
predicted by Eq.~(\ref{tc}) with the exponent $1/(2\gamma)$. This
deviation is possibly due to strong finite-size effects, dominating
the very late stage of the agreement dynamics.

\begin{figure}[t]
\hspace*{-0.50cm}
\includegraphics[width=3.7in]{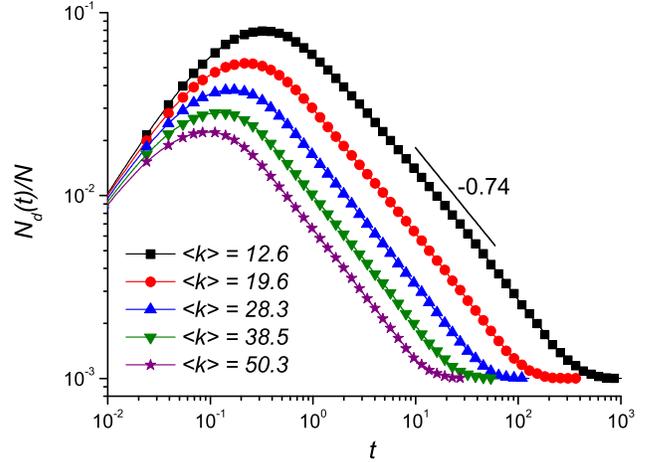}
\vspace*{-1.00truecm}
\caption{(Color online) Time evolution of the (scaled) number of different words
starting from an ``empty" word list initial condition, for various
average degree $\overline{k}$ on log-log scales. The number of nodes
is $N$$=$$1,000$. The straight line segment indicates the asymptotic
power-law decay as determined earlier
[Figs.~\ref{fig.nd-nw-rgg-scaled}(b)], independent of the
neighborhood size $\overline{k}$.} \label{fig.nd-rgg_k}
\end{figure}

For RGGs with many nodes, a relevant control parameter is the
average number of neighbors (or average degree) $\overline{k}$. For
sensor-network-specific implementations, as noted earlier,
$\overline{k}$ can be adjusted by increasing either the density or
the communication range of the nodes. We performed simulations of
the NG for different average neighborhood size $\overline{k}$, as
shown in Fig.~\ref{fig.nd-rgg_k}. The results indicate that the
scaling properties (in terms of $N$) of the time evolution of the
agreement process do not change. The typical convergence times,
however, are significantly reduced by increasing the neighborhood
size. A closer examination of the convergence time reveals that, for
fixed $N$, it scales as $t_c\sim\bar{k}^{-2.6}$, in the
sparse-network limit ($\overline{k}$$\ll$$N$) in two-dimensional
RGGs.

\subsubsection{Agreement-time distributions}

\begin{figure}[t]
\hspace*{-0.50cm}
\includegraphics[width=3.7in]{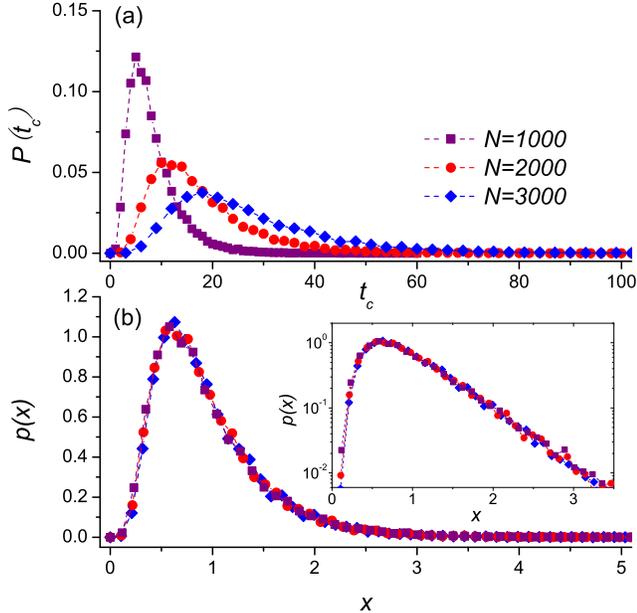}
\vspace*{-1.00truecm}
\caption{(Color online) (a) Probability densities of the convergence time for three
systems sizes. Data are gathered from $100,000$ independent realizations of the NG on RGG.
The average degree of the underlying RGGs is $\overline{k}$$\approx$$50$.
(b) Probability densities for the scaled variable $x=t_c/\langle t_c(N)\rangle$ for the same data.
The inset shows the same scaled histograms on log-lin scales.}
\label{fig.prob-conv}
\end{figure}

In addition to the average agreement (or convergence) time $\langle t_c\rangle$
(time until global agreement is reached), we also measured the
standard deviation $\Delta t_c$ [Fig.~\ref{fig.conv-max-rgg}], and
constructed the probability density (normalized histograms)
$P(t_c,N)$ for $N$ nodes [Fig.~\ref{fig.prob-conv}]. Since in this
subsection, we analyze the full probability density of this
observable, we use brackets for denoting the ensemble-averaged value
of the convergence time, $\langle t_c\rangle$, while $t_c$ alone
denotes the stochastic variable, corresponding to a measurement in a
single realization of the NG.

Up to the system sizes we could simulate, the standard deviation,
within error, scales in the same fashion with the number of nodes as
the average itself, $\Delta t_c$$\sim$$N^{1.10}$
($\overline{k}\approx12$) and $\Delta t_c$$\sim$$N^{1.07}$
($\overline{k}\approx50$) [Fig.~\ref{fig.conv-max-rgg}].
[Suppressing large average convergence times and the corresponding
large standard deviations (through modifying the network
communication topology) will be addressed in the next section.]

Further, the {\em shape} of the histograms, for sufficiently large
systems, remains invariant [Fig.~\ref{fig.prob-conv}]. Thus,
introducing the scaled convergence time $x=t_c/\langle
t_c(N)\rangle$, the corresponding scaled probability densities
$p(x)$ for different system sizes collapse onto the same curve.
[Fig.~\ref{fig.prob-conv}(b)]. The above findings indicates that the
convergence-time distribution for the NG is governed by a {\em
single} scale $\langle t_c\rangle$, hence can be written as
\begin{equation}
P(t_c,N)=\frac{1}{\langle t_c(N)\rangle}p(t_c/\langle
t_c(N)\rangle)\;.
\end{equation}
The distributions exhibit exponential tails for large arguments
[Fig.~\ref{fig.prob-conv}(b) inset], a characteristic feature of
opinion dynamics governed by coarsening \cite{KR2003,Wykes}.

\subsubsection{Cluster-size distributions}

\begin{figure}[t]
\hspace*{-0.50cm}
\includegraphics[width=3.7in]{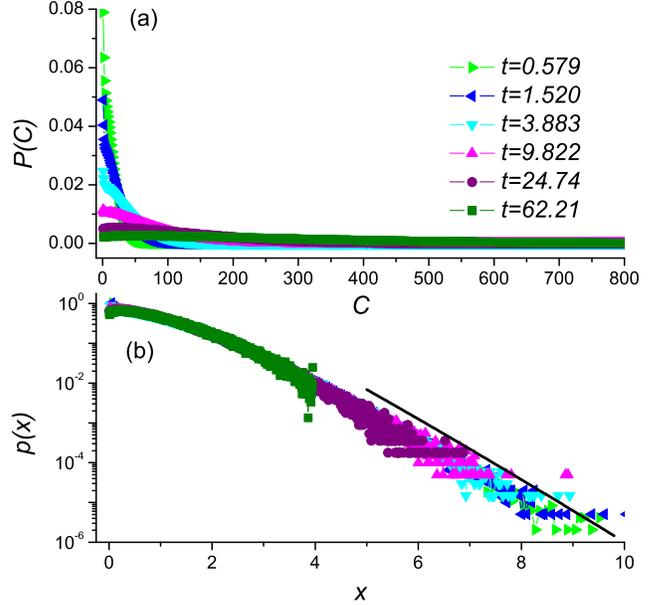}
\vspace*{-1.00truecm}
\caption{(Color online)
(a) Probability densities of the cluster size at different time of
the agreement dynamics. Data are collected through $100,000$ independent
realizations of the NG on RGG. The system size of the underlying RGGs is $N=1000$
and the average degree is $\overline{k}$$\approx$$12$.
(b) Probability densities for the scaled variable
$x=C/\langle C(t)\rangle$ for the same data as in (a) on log-lin scales.
The solis curve represents the best fit exponential-like tail, $\propto\exp(-1.24x^{1.12})$.}
\label{fig.prob-cluster}
\end{figure}

We also studied the probability distribution of the sizes of the
clusters during the agreement dynamics $P(C,t)$ (the normalized
histogram of the sizes of domains with different words at a given
time) [Fig.~\ref{fig.prob-cluster}(a)]. Similar to the previous
subsection, we analyze the full probability density of this
observable, hence we use brackets for denoting the ensemble-averaged
value of the cluster size in the system at time $t$, $\langle
C(t)\rangle$, while $C$ alone denotes the stochastic cluster-size
variable (sampled at an instant $t$ in a single realization of the
NG).

Since the agreement process is governed by coarsening, one expects
that this distribution exhibits dynamic scaling, i.e.,
\begin{equation}
P(C,t)=\frac{1}{\langle C(t)\rangle}p(C/\langle C(t)\rangle)\;.
\end{equation}
Thus, $p(x)$, the distribution of the scaled cluster sizes
$x=C/\langle C(t)\rangle$ remains invariant for different times. Our
simulations confirm this picture, except for very early times
(growth phase with initial domains forming) and for very late times
(where finite-size effects dominate)
[Fig.~\ref{fig.prob-cluster}(b)]. The cluster-size distribution
exhibit exponential-like tails for large arguments
\cite{Derrida1991,Derrida1996,EBN1997,Kumar2001}, as can be seen in
Fig.~\ref{fig.prob-cluster}(b).

\section{Naming Games in Small-World-Connected Random Geometric Networks}

In light of recent results on NG on one-dimensional SW networks
\cite{Baronchelli_2006b}, we now consider accelerating the
agreement process by adding random long-range communication links
between a small fraction of nodes of the RGG. Such networks have
long been known to speed up the spread of local information to
global scales \cite{Watts98,Watts99,NEWMAN_SIAM,KHK_PRL2005}, with
applications ranging from synchronization problems in distributed
computing \cite{KNGTR03a} to alarm-detection schemes in wireless
sensor networks \cite{LKS_INSS2006}. For sensor networks, this can
be implemented either by {\em adding} a small fraction of sensors
equipped with long-range unidirectional antennas (``physical"
long-range connections) or by establishing designated multi-hop
transmission patterns (``logical" long-range connections) between
certain nodes \cite{HELMY_2003}.

We construct the small-world-like RGG (SW RGG) as follows. We
start with the original RGG (embedded in $d$ dimensions, where
$d$$=$$2$ in this paper). Then we {\em add} ``long-range" links (or ``shortcuts")
between randomly chosen nodes in such a way that the
total number of long-range links per node (the density of random
links) is $p$.  This SW construction differs slightly from the
original Watts-Strogatz one \cite{Watts98} (also used by Dall'Asta
et al. \cite{Baronchelli_2006b}), where random links are
introduced by ``rewiring" some of the original connections. The
resulting network, however, has the same universal properties in
the small-$p$, large-$N$ limit \cite{NW_1999}, which is the center
of our interest. Further, it is also motivated by actual
implementations in sensor networks, where long-range
``channels" are established in addition to the existing local ones.

\subsection{Basic Scaling Considerations}

Before presenting simulation results, using scaling arguments, one
can obtain an order of magnitude estimate for the crossover time
$t_{\times}$ present in the SW RGG and for the time to reach
global agreement $t_c$ \cite{Baronchelli_2006b}. In SW networks,
embedded in $d$ dimensions, the typical (Euclidean) distance between nodes with shortcuts
scales as $l_{SW}$$\sim$$p^{-1/d}$ \cite{NW_1999,Barrat1999,Barthelemy1999}.
Starting from empty initial word lists
word, for early times (following the creation of ${\cal
O}(N/(\overline{k}+1))$) different words in the system), the
system will exhibit  coarsening, until the typical linear size of the
growing domains, $\xi(t)$$\sim$$t^{\gamma}$, becomes comparable to
$l_{SW}$. (Here, both lengthscale measures are understood in terms of the underlying Euclidean metric.)
After that time, the agreement process is governed by
the presence of random long-range connections, yielding
mean-field-like behavior. Hence the crossover from $d$-dimensional
coarsening to mean-field-like dynamics occurs when
$t^{\gamma}$$\sim$$p^{-1/d}$, yielding
\begin{equation}
t_\times\sim p^{-1/(d\gamma)}\;. \label{t_x}
\end{equation}
In a system of $N$ agents, the above crossover is only displayed if
the convergence time of the original system with no random links
would exceed the above crossover time
$N^{1/d\gamma}$$\gg$$p^{-1/d\gamma}$, which is equivalent to the
condition for the onset of the SW effect $N$$\gg$$p^{-1}$
\cite{NW_1999,Baronchelli_2006b}. Following the above system-size
independent crossover time, the agreement dynamics is of mean-field
like, and one can expect to observe a scaling behavior closer to
that of FC networks \cite{Baronchelli_2005a}. In particular, the
time to reach global agreement is expected to scale as
\cite{Baronchelli_2006b}
\begin{equation}
t_c\sim N^{1/2}
\label{t_cmf}\;,
\end{equation}
a significant reduction compared to that of the ``pure" RGG with no
long-range links where $t_c$$\sim$$N^{1.1}$.

\subsection{Simulation Results}

\begin{figure}[t]
\hspace*{-0.50cm}
\includegraphics[width=3.7in]{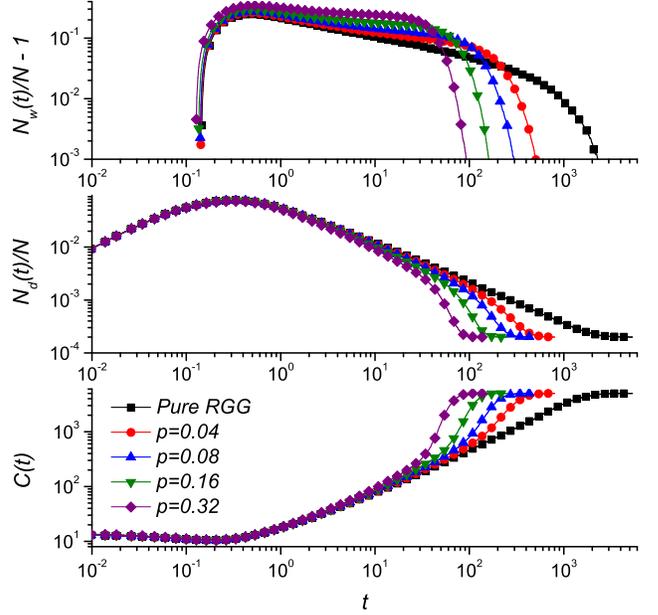}
\vspace*{-1.00truecm}
\caption{(Color online) Time evolution of the (scaled) (a) total number of words, (b)
number of different words and (c) the average cluster size for SW RGGs
on log-log scales, starting from an
``empty" word list initial condition, for various density of
long-range links $p$, averaged over $1,000$ independent
realizations of the NG on RGG. The number of nodes is $N$$=$$5,000$ with
average degree $\overline{k}$$\approx$$12$.}
\label{fig.nd-nw-c-swrgg}
\end{figure}

Simulating the NG on SW RGGs qualitatively confirms the above
scaling scenario. Following the very early-time development of
${\cal O}(N/(\overline{k}+1))$ different words, the system of
SW-networked agents, exhibits slow coarsening, with only small
corrections to the behavior of the pure RGG
[Fig.~\ref{fig.nd-nw-c-swrgg}]. In fact, this early-time coarsening
on SW RGGs is slightly slower compared to pure RGGs due to the
effective pinning of interfaces near the shortcuts
\cite{Baronchelli_2006b,Boyer2003,Castellano2003,Castellano2005}. In
the NG on SW networks, however, the agreement process only slows
down \cite{Baronchelli_2006b}, but is not halted by ``frozen"
(metastable) disordered configurations
\cite{Boyer2003,Castellano2005}. After a $p$-dependent crossover
time [Eq.~(\ref{t_x})], (when the typical size of the growing
clusters becomes comparable to the SW length scale), an exponential
convergence begins to govern the agreement process. This final-stage
fast approach toward consensus sets in earlier for increasing values
of the density of shortcuts $p$, yielding a significantly reduced
convergence time compared to that of the NG on the ``pure" RGG. The
temporal behavior of the relevant observables for various values of
$p$ can be observed in Fig.~\ref{fig.nd-nw-c-swrgg}.

\begin{figure}[t]
\hspace*{-0.50cm}
\includegraphics[width=3.7in]{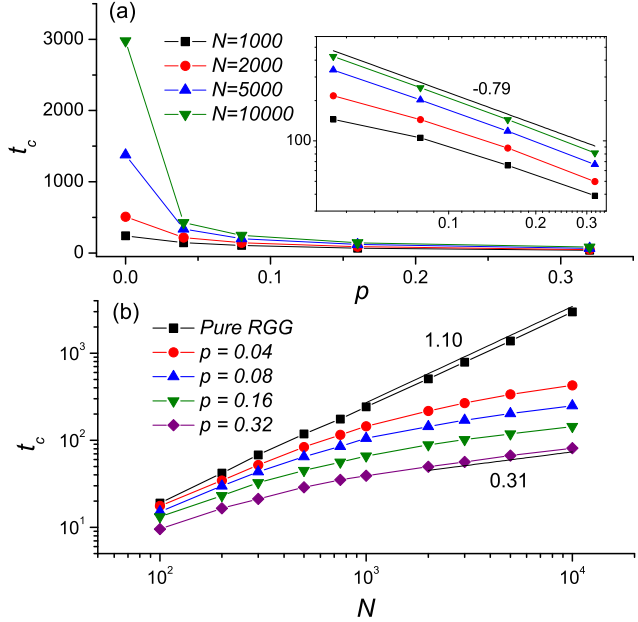}
\vspace*{-1.00truecm}
\caption{(Color online) Average  convergence time $t_c$ for
SW RGGs. (a) as a function of the density of shortcuts for various
system sizes. The inset shows the same data on
log-log scales. The straight lines corresponds to an estimate of
the associated (asymptotic) power-law.
(b) as a function of the number of nodes on log-log
scales for various density of long-range links $p$. The curves shown are obtained by averaging over
$1,000$ independent realizations of of the NG on RGG. The average degree of
the underlying RGGs is $\overline{k}$$\approx$$12$.}
\label{fig.conv-p-swrgg}
\end{figure}

Plotting the convergence time vs the density of long-range links, as
shown in Fig.~\ref{fig.conv-p-swrgg}(a), suggests that (for
sufficiently large but {\em fixed} $N$) the convergence time
approaches an asymptotic power-law $t_{c}$$\sim$$p^{-s}$ with
$s$$=$$0.79$$\pm$$0.01$ \cite{Baronchelli_2006b}.
On the other hand, for fixed $p$ and increasing $N$,
the convergence time increases with $N$,
$t_c$$\sim$$N^{\alpha_{\rm SW}}$, with $\alpha_{\rm
SW}$$=$$0.31\pm0.01$ [Fig.~\ref{fig.conv-p-swrgg}(b)].
The agreement process is much faster than on a two-dimensional
regular grid or RGG and is closer to the anticipated mean-field-like
behavior [Eq.~(\ref{t_cmf})] \cite{Baronchelli_2006b}. Thus, in the
small-world regime ($Np$$\gg$$1$) the convergence time depends on
both the system size and density of random links,
$t_c$$\sim$$N^{\alpha_{\rm SW}}/p^s$.

\subsubsection{Finite-size scaling for the agreement time on SW-RGGs}

\begin{figure}[t]
\hspace*{-0.50cm}
\includegraphics[width=3.7in]{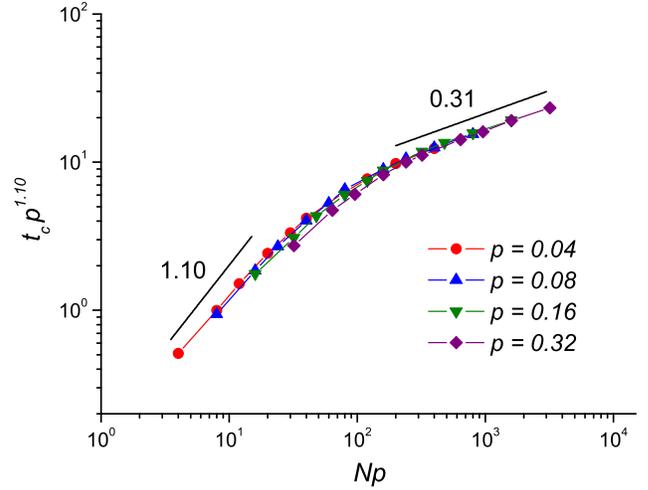}
\vspace*{-1.00truecm}
\caption{(Color online) Scaled plot of data shown in Fig.~\ref{fig.conv-p-swrgg},
as suggested by the finite-size scaling argument [Eq.~(\ref{tcswscale})].
The straight line segments correspond to the best-fit (asymptotic)
power-law behavior of the scaling function $g(x)$ with exponents $1.10$
and $0.31$, for small and large arguments, respectively, as described in the text [Eq.~(\ref{gx})].}
\label{fig.conv-N-swrgg}
\end{figure}

In the pure-RGG limit ($Np$$\ll$$1$), $t_c$ only depends on $N$,
$t_c$$\sim$$N^{\alpha_{\rm RGG}}$ with $\alpha_{\rm
RGG}$$\approx$$1.10$ [Fig.~\ref{fig.conv-p-swrgg}(b)] (since,
essentially there are no shortcuts in the system). On the other
hand, as seen above, in the SW-regime ($Np$$\gg$$1$), the agreement
time scales as $t_c$$\sim$$N^{\alpha_{\rm SW}}/p^s$. One then can
construct the full scaling behavior of $t_c(p,N)$, capturing the
above two finite-size behaviors as limiting cases on SW-connected
RGGs,
\begin{equation}
t_c(p,N)\sim \frac{N^{\alpha_{\rm SW}}}{p^s}f(Np)\;, \label{tcsw}
\end{equation}
where $f(x)$ is a scaling function such that
\begin{equation}
f(x)\sim\left\{
    \begin{array}{ll}
        x^s & \mbox{if $x$$\ll$$1$} \\
        {\rm const.} & \mbox{if $x$$\gg$$1$}
    \end{array}
\right. \;.
\end{equation}
The pure RGG limit ($Np$$\ll$$1$) is recovered, provided that
$t_c$$\sim$$(N^{\alpha_{\rm SW}}/p^s)(Np)^s$$\sim$$N^{\alpha_{\rm SW}+s}$$\sim$$N^{\alpha_{\rm RGG}}$,
i.e.,
\begin{equation}
\alpha_{\rm RGG} = \alpha_{\rm SW}+s\;.
\label{alpha2d}
\end{equation}
Our measured ``phenomenological" exponents $\alpha_{\rm
RGG}$$\approx$$1.10$, $\alpha_{\rm SW}$$\approx$$0.31$, and
$s$$\approx$$0.79$, satisfy the above proposed asymptotic scaling
relation. For analyzing our data, Eq.~(\ref{tcsw}) can also be
rewritten as
\begin{equation}
t_c(p,N)\sim \frac{(Np)^{\alpha_{\rm SW}}}{p^{s + \alpha_{\rm
SW}}}f(Np) \sim \frac{1}{p^{\alpha_{\rm RGG}}}g(Np)\;,
\label{tcswscale}
\end{equation}
where $g(x)=x^{\alpha_{\rm SW}}f(x)$. Thus, plotting $t_c
p^{\alpha_{\rm RGG}}$ vs $Np$ should yield data collapse, together
with the asymptotic small- and large-argument exponents of $g(x)$,
$\alpha_{\rm RGG}$ and $\alpha_{\rm SW}$, respectively
[Fig.~\ref{fig.conv-N-swrgg}],
\begin{equation}
g(x)\sim\left\{
    \begin{array}{ll}
        x^{\alpha_{\rm RGG}} & \mbox{if $x$$\ll$$1$} \\
        x^{\alpha_{\rm SW}} & \mbox{if $x$$\gg$$1$}
    \end{array}
\right. \;.
\label{gx}
\end{equation}

\section{Summary and Outlook}

In this paper, we studied a prototypical agent-based model, the
Naming Game, on Random Geometric Graphs and SW-connected RGGs
embedded in two dimensions. While the underlying RGG communication
topology is motivated by large-scale sensor networks, the NG on
these networks captures fundamental features of agreement dynamics
of spatially-embedded networked agent-based systems. We have found
that, qualitatively similar to two-dimensional regular networks
\cite{Baronchelli_2006a}, the NG on RGG can be reasonably well
described by the physical theory of coarsening. In particular, local
clusters of nodes sharing the same word quickly form, followed by
slow coarsening of these clusters in the late stage of the dynamics.
The typical length scale grows as $\xi(t)$$\sim$$t^{\gamma}$ with
the coarsening exponent estimated to be $0.35$$<$$\gamma$$<$$0.40$.
Our simulation results also indicate that the average time to reach
global agreement is of ${\cal O}(N^{1.08\pm0.03})$ (for {\em fixed}
average degree). The above results imply that, at least for the
range of finite system sizes studied here (up to $N=5,000$), the
characteristic length scale in two-dimensional RGGs grows slower
than $1/2$. This deviation, in part, may very well be attributed to
the effectively small system sizes that we could study. Similarly
strong transients and finite-size corrections, due to the presence
of ``soft domain walls" and ``vertices" (domain-wall intersections),
also made the precise determination of the asymptotic coarsening
exponent difficult in the two-dimensional large-$Q$ (effectively
$Q$$=$$\infty$) Potts models
\cite{Glazier1990,Grest1983a,Grest1983b,Kaski1985,Kaski1987,Grant1990}.
On the the other hand, based on our Monte Carlo studies, we cannot
rule out the possibility that the deviation from the
$\gamma$$=$$1/2$ coarsening exponent is the result of the inherent
{\em local random} random structure of RGGs (in contrast to
regular two-dimensional grids \cite{Baronchelli_2006a}).

While in this paper we did not address the message complexity of the
NG explicitly, one can make an order of magnitude estimate for the
typical number of messages needed to reach global agreement on RGGs
for an efficient implementation. (In sensor networks, this quantity
is also relevant since it corresponds to the global energy
consumption.) Once the coarsening process begins, nodes inside the
clusters have reached agreement with all their neighbors, of which
they are readily aware, hence, they no longer have to initiate
broadcasts any longer. Thus, only these ``active" nodes, found at
the interfaces between these cluster (which have at least one
neighbor with different words), will initiate broadcast for word
matching. Using that the number of nodes at the interfaces scales as
$N/t^{\gamma}$ [Eq.~(\ref{Nw})], and integrating this
expression up to $t_c$$\sim$$N^{1.08\pm0.03}$, one finds that the total number of messages
needed to be exchanged until global agreement is reached is of
${\cal O}(N^{1.68\pm0.05})$.

In an attempt to accelerate the agreement process by changing the
communication topology between agents, we also studied the
SW-connected version of the two-dimensional RGG. By adding a small
density of shortcuts ``on top" of the RGG, resulting in a SW-like
network, the convergence time is strongly reduced and becomes of
${\cal O}(N^{0.31})$, similar to the behavior of NG on the
Watts-Strogatz SW network \cite{Baronchelli_2006b}.

In future works we will investigate the NG on more realistic
communication topologies, motivated by and relevant to wireless
sensor networks, in particular, random spatial networks with
heterogeneous range distribution, minimum-node-degree networks
\cite{Bassler2007}, and also networks with dynamically changing
connectivities.

\acknowledgments We thank B. Yener, J.W. Branch, and A. Barrat for
comments on this work. This research was supported in part by NSF
Grant Nos.\ DMR-0426488 (G.K. and Q.L.), NGS-0103708 (B.K.S. and
Q.L.), and by Rensselaer's Seed Program. B.K.S. was also supported
through participation in the International Technology Alliance
sponsored by the U.S. Army Research Laboratory and the U.K. Ministry
of Defence under Agreement Number W911NF-06-3-0001.


\end{document}